# Studies on the proximity effect in Bi-based high-temperature superconductor/manganite heterostructures

Gayathri V[1,2], E P Amaladass[1,2], A T Sathyanarayana[1,2], T Geetha Kumary[1], R. Pandian[1,2], Pooja Gupta[2,3], Sanjay K Rai[3], and Awadhesh Mani[1,2*]

[1]*Materials Science Group, Indira Gandhi Centre for Atomic Research, Kalpakkam 603102, India*

[2]*Homi Bhabha National Institute, Training School Complex, Anushaktinagar, Mumbai 400094, India*

[3]*Accelerator Physics and Synchrotrons Utilisation Division, Raja Ramanna Centre for Advanced Technology, Indore 452013, India*

*Corresponding author: mani@igcar.gov.in; geetha@igcar.gov.in*

**Abstract**

The effect of proximity of the magnetism of the Pr-based manganite ($Pr_{0.6}Sr_{0.4}MnO_3$) on the superconductivity of Bi-based high-temperature superconductor ($Bi_{1.75}Pb_{0.25}Sr_2Ca_2Cu_3O_{10+\delta}$) was studied based on the results obtained from the magnetotransport and magnetization measurements. Decrease in the values of the upper critical field ($H_{C2}(0)$) and an increase in the width of the superconducting transition ($\Delta T_C$) of $Bi_{1.75}Pb_{0.25}Sr_2Ca_2Cu_3O_{10+\delta}$ were observed in proximity with the manganite. The combined effect of magnetic exchange interaction arising from the manganite, the leakage of Cooper-pairs from the superconductor into the manganite, and the diffusion and transport of spin-polarized electrons from the manganite into the superconductor were found to modify the superconducting properties of $Bi_{1.75}Pb_{0.25}Sr_2Ca_2Cu_3O_{10+\delta}$. The stacking sequence of the individual layers in these heterostructures was found to dictate the ground state properties of the heterostructure. As a consequence of the proximity effect, the colossal-magnetoresistance (CMR) ratio as high as ~ 99 % observed in the heterostructure makes the thin film heterostructures promising candidates for potential technological applications.

**Introduction**

The proximity effect is a manifestation of the modification in the superconducting properties of a superconductor when in direct contact with a normal metal, another superconductor, or a ferromagnetic metal [1–3]. The overlapping of the wavefunctions creates a gradient in the density of Cooper-pairs. The density of Cooper-pairs increases from the interface towards the bulk. The $T_C$ of the superconductor decreases as a result of the leakage of Cooper-pairs into the normal metal with the simultaneous leakage of some quasiparticles into the superconductor at the interface of the normal metal/superconductor [4]. Here, the thicknesses of the individual layers (thickness of superconductor = $d_{SC}$, thickness of normal metal = $d_N$) and the characteristic coherence length of the superconductor and the normal metal ($\xi_{SC}$ and $\xi_N$, respectively) control the magnitude of the $T_C$ suppression. These coherence lengths ($\xi_{SC}$ and $\xi_N$) are generally of the order of 1 Å to 10 Å in high-temperature superconductors (HTSC). For $d_{SC} \gg \xi_{SC}$, with an increasing value of $d_N$, the value of $T_C$ drops and has a finite limiting value as $d_N$ becomes greater than $\xi_N$. For $d_{SC} \ll \xi_{SC}$, a drastic suppression of $T_C$ is observed, which becomes experimentally undetectable as $d_N \sim \xi_N$. When two superconductors having different $T_C$s ($T_{C1}$ and $T_{C2}$) are brought in intimate contact with each other, as a consequence of the proximity effect, an intermediate $T_C$ is achieved [5–7]. In the case of a ferromagnetic metal/superconductor configuration, the magnetic exchange energy of the ferromagnet tends to align the spins of the Cooper-pairs in the same direction. The interaction of the spin-polarized quasiparticles from the ferromagnet with the Cooper-pairs of the superconductor creates a Zeeman splitting of the electronic energy levels. If the Zeeman energy exceeds the superconducting energy gap, superconductivity will be suppressed. Due to the proximity effect, the Cooper-pair wavefunction can extend from the superconductor into the ferromagnet resulting in damped oscillatory behaviours like non-monotonic variation of $T_C$ with $d_N$ and spatial oscillation of the electron density of states [8–11].

Research on the proximity effect between superconductivity and magnetism deepen our understanding of the fundamental physics and these effects can open up tantalizing possibilities for prospective technological applications [1,12]. The literature is rich with studies of the proximity effect between low-$T_C$ superconductors and ferromagnets. A few notable studies among them were on Gd/Nb nanostructures by Jiang et al. [8], bilayers of Pb/Ni by Bourgeois and Dynes [9], structures of Fe/Nb/Fe by Muhge et al. [10], and V/Fe systems by Aarts et al. [11], exhibiting different proximity behaviours depending on their transmittance at the superconductor/ferromagnet interface. For the past few years, there has been a lot of research on the coexistence of magnetism with HTSC with the major focus on $YBa_2Cu_3O_7$ [13–17]. These studies have resulted in several exotic phenomena like the formation of π-junctions, spatial modulation of the order parameter, domain-wall superconductivity, etc. [2,3,18,19] at the $YBa_2Cu_3O_7$/ferromagnet interface. However, Bi-based HTSC are less explored for their proximity effects with the ferromagnets [20–24]. Furthermore, the stacking sequence of the individual layers in these heterostructures can also dictate their ground state properties [25], making the study all the more exciting.

**Experiment**

The heterostructures of $Bi_{1.75}Pb_{0.25}Sr_2Ca_2Cu_3O_{10+\delta}$ (B(Pb)SCCO) and $Pr_{0.6}Sr_{0.4}MnO_3$ (PSMO) in two different stacking sequences were prepared on single crystalline *(1 0 0)* oriented substrate of $SrTiO_3$ (STO) by using pulsed laser deposition technique. The first heterostructure abbreviated as B(Pb)SCCO/PSMO/STO is fabricated by initially depositing PSMO magnetic layer of 30 nm thickness on STO. Subsequently, a 300 nm superconducting layer of B(Pb)SCCO was deposited on about half of the PSMO surface. The heterostructure is referred to as B(Pb)SCCO/PSMO/STO in the following discussions. The second heterostructure of $Bi_{1.75}Pb_{0.25}Sr_2Ca_2Cu_3O_{10+\delta}$ (B(Pb)SCCO) and $Pr_{0.6}Sr_{0.4}MnO_3$ (PSMO) thin film in the reverse stacking sequence, abbreviated as PSMO/B(Pb)SCCO/STO, is fabricated by initially depositing B(Pb)SCCO layer of ~ 300 nm thickness on STO. Subsequently, a ~ 30 nm magnetic layer of PSMO was deposited on about half of the B(Pb)SCCO surface. The fabricated heterostructure in the reverse stacking sequence is referred to as PSMO/B(Pb)SCCO/STO in the following discussions. The films were deposited under optimized deposition conditions using the PLD technique. The respective targets for the film deposition was synthesized by the solid-state reaction route [7,24,26–31].

The grown heterostructures were characterized for their structural properties using engineering applications beamline (BL-02), Indus-2 Synchrotron facility, RRCAT, Indore (India) [32]. X-rays of wavelength 0.826 Å were utilised for the powder X-ray diffraction (XRD) and high-resolution X-ray diffraction (HR-XRD) measurements. Data was collected in the reflection geometry using silicon strip detector (MYTHEN2 X 1K) from Dectris. Scanning electron microscopy (SEM) measurements were performed to study the morphology using a field emission scanning electron microscope (FE-SEM) with 30 kV acceleration voltage, model SUPRA 55 by Carl Zeiss, Germany. The energy-dispersive X-ray spectroscopy (EDS) measurements were performed using a 10 mm$^2$ liquid nitrogen-free Silicon drift detector (SDD) by Oxford Instruments Inc. (model: X-act) attached to the SEM for EDS analysis (by INCA EDS software) of the fabricated heterostructure. The fracture cross-sectional SEM imaging of the sample was performed to confirm the thickness of the grown heterostructure. Magnetization measurements were performed in the temperature range of 4 K to 300 K and in the applied magnetic field range of 0 T to ± 7 T, applied parallel (H||*ab*) to the plane of the heterostructure, using a Quantum Design Ever-Cool SQUID magnetometer. The substrate contribution towards the magnetization data was eliminated by repeating the measurements in identical conditions on a bare substrate of the same dimension and subtracting it from the obtained data of the heterostructure. Magnetotransport measurements were performed in the linear geometry, using a 15 T cryofree MR system from

Cryogenic, UK, at temperature from 4 K to 300 K and in the magnetic field range of 0 T to ± 15 T, applied both parallel (H||*ab*) and perpendicular (H||*c*) to the plane of the heterostructure. The special measurement protocol of cooling and heating in unequal fields (CHUF) was performed to study the behaviour of the kinetically arrested phases in these heterostructures.

Figure 1 illustrates the schematic diagram of the B(Pb)SCCO/PSMO/STO heterostructure. During the transport measurements, the current was passed across the whole sample through leads 1 and 8. The voltage drop was determined across three regimes: (a) $V_1$ measured between leads 2 and 3 (on the B(Pb)SCCO/PSMO/STO heterostructure), (b) $V_2$ measured between leads 4 and 5 (junction between PSMO and B(Pb)SCCO/PSMO/STO heterostructure), and (c) $V_3$ measured between leads 6 and 7 (on the PSMO layer). This configuration, as shown in Figure 1, allows us to explore the different parts of the sample simultaneously.

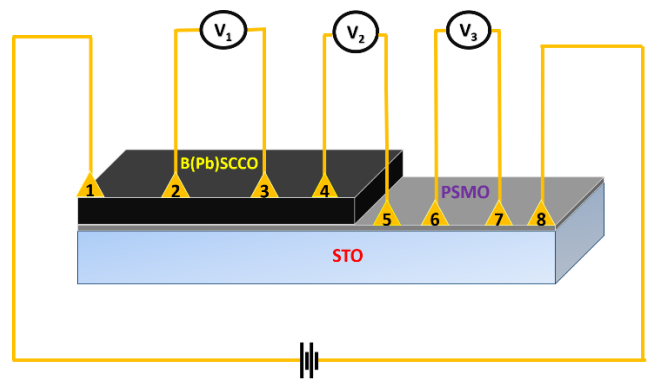

*Figure 1. Schematic representation of the B(Pb)SCCO/PSMO/STO heterostructure thin film used for magnetotransport measurement*

Figure 2 illustrates the schematic diagram of the PSMO/B(Pb)SCCO/STO heterostructure. During the transport measurements, the current was passed across the whole sample through leads 1 and 8. The voltage drop was determined across the three regimes: (a) $V_1$ measured between leads 2 and 3 (on the B(Pb)SCCO layer), (b) $V_2$ measured between leads 4 and 5 (across the junction between B(Pb)SCCO and the PSMO/B(Pb)SCCO/STO heterostructure), and (c) $V_3$ measured between leads 6 and 7 (on the PSMO/B(Pb)SCCO/STO heterostructure).

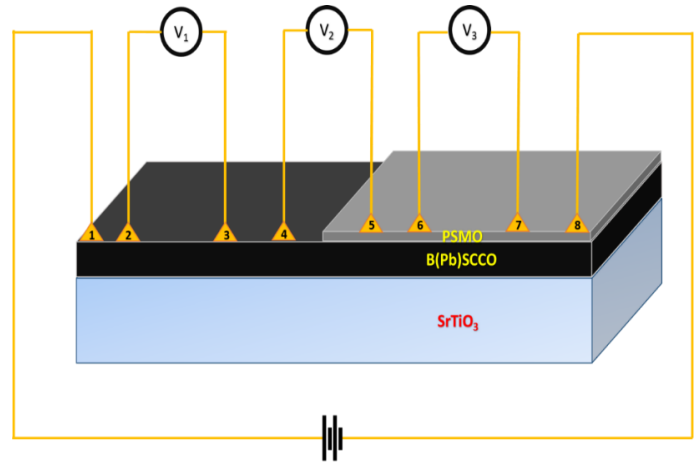

*Figure 2. Schematic representation of the PSMO/B(Pb)SCCO/STO heterostructure thin film used for magnetotransport measurement*

**Results and discussion**

***B(Pb)SCCO/PSMO/STO***

Structural properties

Figure 3 and the corresponding inset represent the XRD pattern obtained for the fabricated B(Pb)SCCO/PSMO/STO heterostructure in the powder X-ray diffraction (XRD) mode and the high-resolution X-ray diffraction (HR-XRD) mode, respectively. The HR-XRD is performed near the *(2 0 0)* reflection of STO (located at 24.4°) in the 2θ range of 22° to 27°. The HR-XRD pattern obtained for the PSMO/STO single layer is also shown in the inset of Figure 3 for comparison. The peak positions were indexed by comparing them with the ICDD data [33–35]. The presence of $Bi_{1.75}Pb_{0.25}Sr_2Ca_2Cu_3O_{10+\delta}$ and $Pr_{0.6}Sr_{0.4}MnO_3$ layers in the heterostructure was confirmed from the XRD pattern. The magnetic PSMO layer crystallized in the orthorhombic structure with the space group *Pnma* and the superconducting B(Pb)SCCO layer is stabilized in the tetragonal structure with the space group *I4/mmm*.

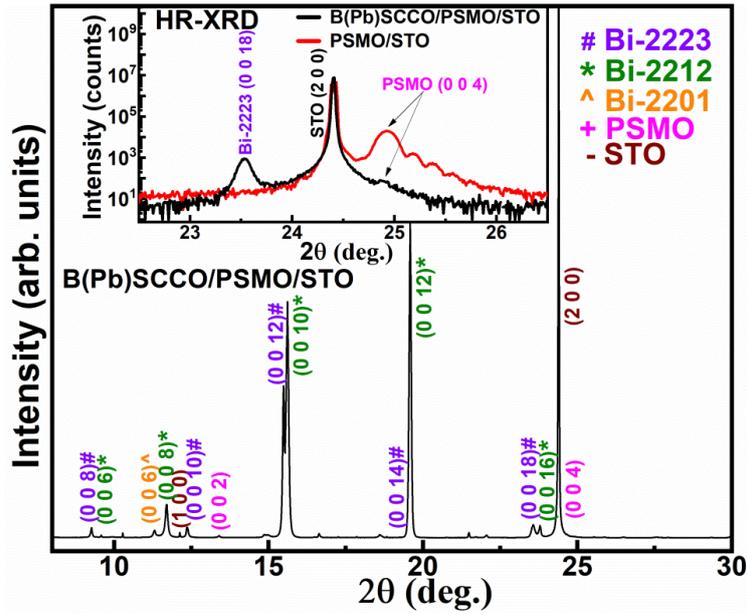

*Figure 3. XRD pattern of the B(Pb)SCCO/PSMO/STO heterostructure thin film. Inset shows the HR-XRD of B(Pb)SCCO/PSMO/STO heterostructure and PSMO/STO single layer near (2 0 0) reflection of STO.*

The presence of *(0 0 2l)* peaks of PSMO and B(Pb)SCCO confirm the *c*-axis preferential orientation of the heterostructure. A peak pertaining to the PSMO layer of the heterostructure was obtained at 24.9° near the *(2 0 0)* reflection of STO, as shown in the inset of Figure 3. The *c*-lattice parameter of PSMO, as determined from the *c*-axis oriented peak positions using Bragg's law, was found to be ~ 7.65 Å and ~ 7.66 Å in the single layer and the heterostructure, respectively. The calculated *c*-lattice parameters are comparable to the corresponding PSMO bulk value of ~ 7.67 Å. The *c*-lattice parameters of the coexisting BSCCO phases (Bi-2223, Bi-2212, and Bi-2201) are also very close to the corresponding values of their single-layer and bulk counterparts.

Morphological analyses

The thickness and the interface of the grown B(Pb)SCCO/PSMO/STO heterostructure sample were revealed using the fracture cross-sectional view mode of SEM. Figure 4 represents the cross-sectional SEM micrographs of the B(Pb)SCCO/PSMO/STO heterostructure. The micrographs show the presence of ~ 300 nm of layered B(Pb)SCCO deposited on top of a ~ 30 nm layer of PSMO grown on the STO substrate with sharp interfaces. Such high-quality, sharp interfaces open the possibility for nano-device fabrication based on these bilayer heterostructures. The presence of all the elements in the heterostructure was confirmed using energy-dispersive X-ray spectroscopy (EDS).

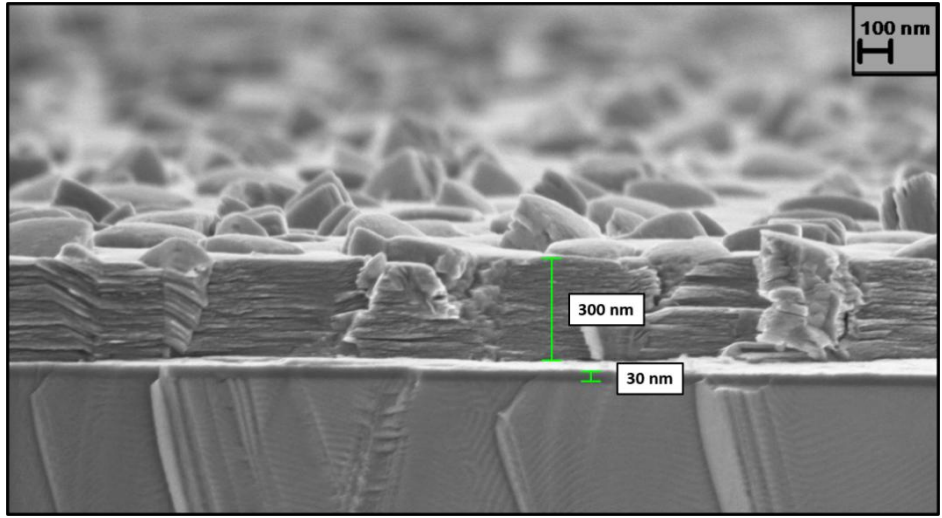

*Figure 4. Cross-sectional SEM of the B(Pb)SCCO/PSMO/STO heterostructure illustrating the stacking of the individual layers and their thicknesses.*

Magnetotransport and magnetization measurements

Figure 5 (a) shows the temperature dependent resistance curve (R(T)) as a function of magnetic field obtained for the B(Pb)SCCO/PSMO/STO heterostructure in the first regime, i.e., on the bilayer side of the B(Pb)SCCO/PSMO/STO heterostructure, where $V_1$ is measured between leads 2 and 3, in the H∥$c$ configuration. Inset in figure 5 (a) shows the field dependent R(T) curve obtained for the B(Pb)SCCO/STO single-layer thin film. Figure 5 (b) shows the field dependent R(T) curve obtained for the B(Pb)SCCO/PSMO/STO heterostructure in the third regime, i.e., on the PSMO layer, where $V_3$ is measured between leads 6 and 7. Inset in figure 5 (b) represents the R(T) curve of the 30 nm annealed PSMO/STO single layer.

Since the 30 nm annealed PSMO single layer is in the ferromagnetic metallic state below ~ 170 K and the 300 nm annealed B(Pb)SCCO single layer is superconducting below

110 K [36], as shown in the insets of figure 5 (a) and (b), respectively, three factors need to be considered in the study of the proximity effect between B(Pb)SCCO and PSMO at temperatures below the $T_C$ (110 K) of B(Pb)SCCO:

(a) Magnetic exchange interaction arising from the ferromagnetic PSMO

(b) Leakage of Cooper-pairs from the superconducting B(Pb)SCCO into the ferromagnetic PSMO

(c) Diffusion and transport of spin-polarized electrons from the ferromagnetic PSMO into superconducting B(Pb)SCCO

A metallic trend is observed in the normal state on the bilayer side of the B(Pb)SCCO/PSMO/STO heterostructure, as shown in figure 5 (a). Two superconducting transitions are observed with decreasing temperature, with $T_{C,ON1}$ and $T_{C,ON2}$ at 90 K and 50 K, respectively. $T_{C,ON1}$ and $T_{C,ON2}$ can be attributed to the superconducting drops corresponding to the Bi-2223 and Bi-2212 phases. The observed $T_C$s are much lower than the B(Pb)SCCO/STO thin film [28], as shown in figure 5 (a) and its inset. This can be due to the competition between the two opposing properties leading to the detrimental impact of the ferromagnetic PSMO on the superconductivity of B(Pb)SCCO. For H = 0 T, $T_{C,OFF}$ is obtained at 10 K. With increasing H value, the resistive curve exhibits a diminished superconducting property accompanied by the fanning out of the resistive curve leading to a wide transition width. The observed decrease in $T_C$s of both phases with respect to their values obtained in the single-layer film and the associated broadening of the transition width indicates the proximity-induced effect of the FM PSMO layer.

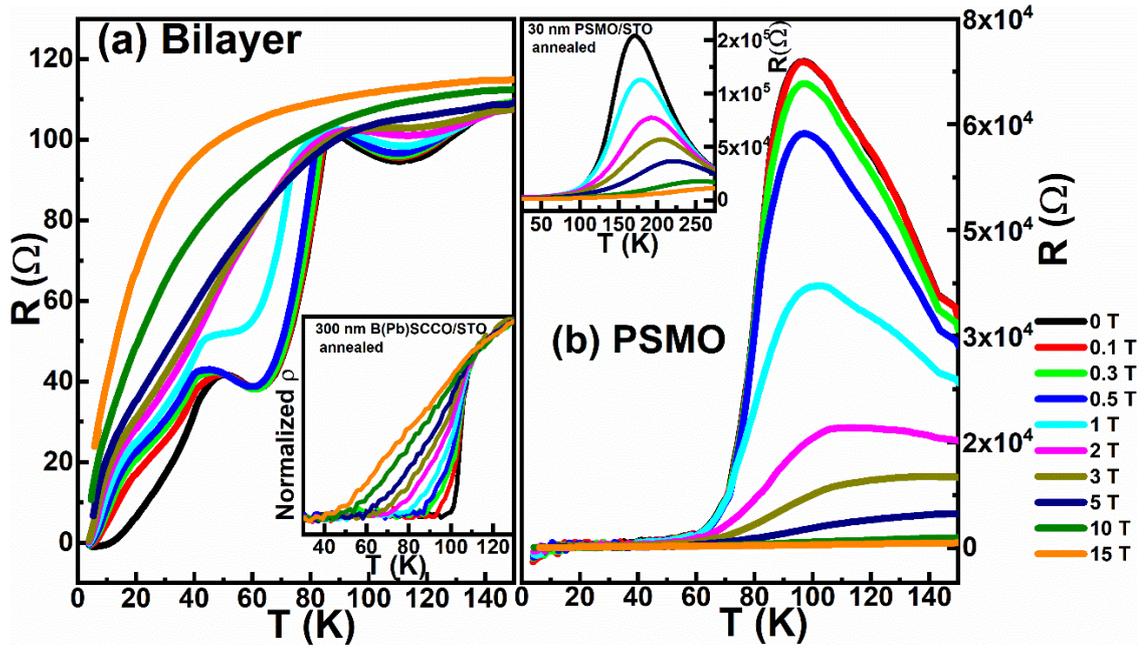

*Figure 5. Magnetic field-dependent normalized R(T) curves of B(Pb)SCCO/PSMO/STO heterostructure for H||c measured across (a) leads 2 and 3 (the heterostructure), and (b) leads 6 and 7 (PSMO layer). Insets in (a) and (b) represents the magnetic-field dependent R(T) curve of B(Pb)SCCO/STO, and PSMO/STO annealed single-layer thin films, respectively.*

On the PSMO regime of the B(Pb)SCCO/PSMO/STO heterostructure, a metal-insulator transition is observed at $T_{MIT}$ ~ 98 K. Additionally, with the application of H, the $T_{MIT}$ is found to shift to higher temperatures, and a huge CMR is observed in this regime. It is to be noted that the value of $T_{MIT}$ in the case of the PSMO/STO single layer was observed to be at 170 K, as shown in the inset of figure 5 (b) [36]. Hence, the value of $T_{MIT}$ on the PSMO layer in proximity with B(Pb)SCCO is found to be tuned towards the $T_C$ of B(Pb)SCCO. Figure 6 illustrates the magnetic field response of the resistive curve in MR ratio (%) calculated using equation (1) for the PSMO layer at various temperatures. Here, R(0) and R(H) are the film's resistance obtained at 0 T and an applied magnetic field 'H', respectively. A huge low-field CMR ratio of nearly 99 % is obtained on the PSMO layer for T ≤ 100 K. The magnetic field response of the resistive curve on the PSMO layer is very quick, as seen in Figure 6. The MR ratio (%) switches between 0 % and 99 % within an applied field of 2 T at 20 K. This fast switching behaviour of the MR ratio (%) with H can be used for low-field CMR-based spintronic applications.

$$MR\ ratio\ (\%) = \frac{(R(H) - R(0))}{R(0)} \times 100 \qquad (1)$$

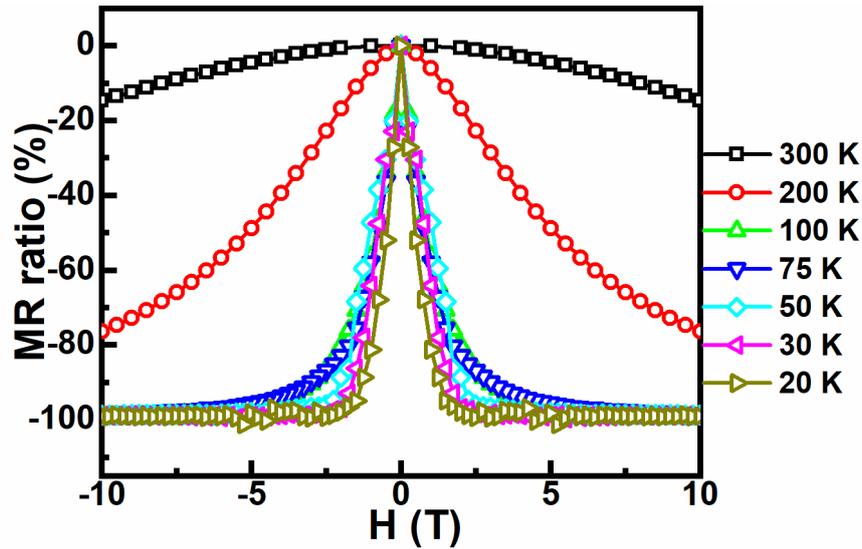

*Figure 6. MR ratio (%) obtained from magnetotransport measurement for the B(Pb)SCCO/PSMO/STO heterostructure at different temperatures in the magnetic field range of ± 10 T in H//ab configuration on the PSMO layer.*

Figure 7 (a) shows the field dependent R(T) curve obtained for the B(Pb)SCCO/PSMO/STO heterostructure in the second regime, i.e., across the junction between the B(Pb)SCCO/PSMO/STO heterostructure and the PSMO layer, where $V_2$ is measured between leads 4 and 5. Across the junction, an insulating behaviour is observed from room temperature down to ~ 70 K, followed by a transition into a metallic state at $T_{MIT}$ ~ 70 K. On further cooling to temperatures below $T_{MIT}$, at ~ 40 K, the resistance further increases with a resistive upturn in the R(T) curve, which may indicate the presence of the scattering centres. With the application of H, an overall decrease in the resistance is observed, thereby exhibiting a huge CMR. This shows that the magnetic exchange energy is dominant over the superconducting condensation energy in this regime. The magnetic field response of the resistive curve in MR ratio (%) calculated using equation (1) for this regime at various temperatures is presented in the inset of figure 7 (a). A huge CMR ratio of nearly 99 % is obtained across the junction for T ≤ 75 K at higher fields. The huge CMR effect exhibited in this regime at higher fields validates that the exchange field arising from the spin-polarized ferromagnetic PSMO layer prohibits the superconducting pairing and aligns the magnetic scattering centres coexisting at low temperatures along the applied magnetic field direction, thereby leading to a negative CMR.

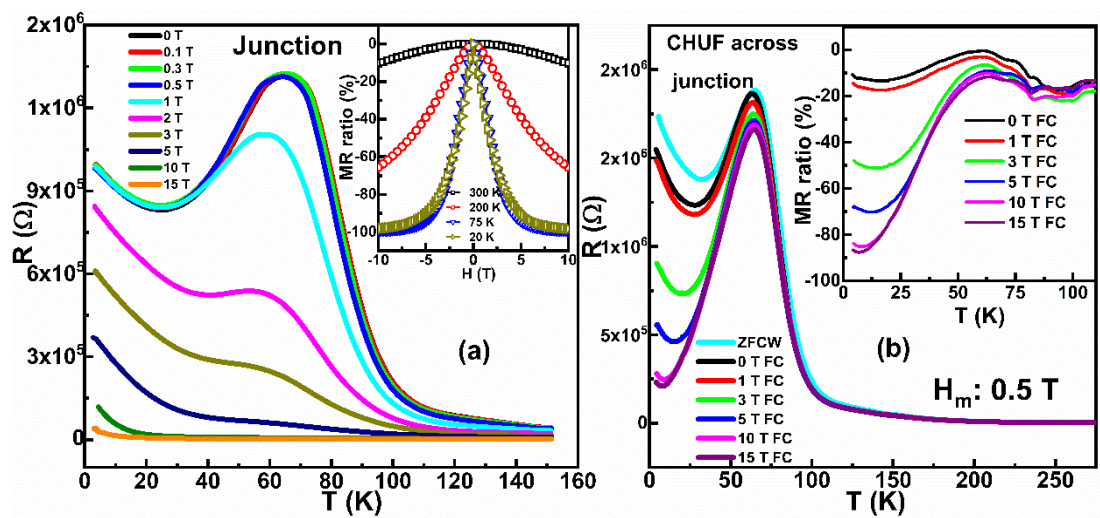

*Figure 7. (a) Magnetic field-dependent R(T) curves obtained for B(Pb)SCCO/PSMO/STO heterostructure for H//c measured across leads 4 and 5 (the junction). The inset in (a) shows MR % obtained from magnetotransport measurement for the junction region of the heterostructure at different temperatures in the magnetic field range of ± 10 T. (b) Cooling and heating in unequal field measurement protocol for B(Pb)SCCO/PSMO/STO heterostructure for $H_m$=0.5 T in the junction region of the heterostructure. The inset in (b) shows the MR ratio (%) of the kinetically arrested FM phase across the junction in the heterostructure*

Cooling and heating in unequal fields (CHUF) measurement across the junction

Cooling and heating in unequal fields (CHUF) is a unique measurement protocol used to study the kinetically arrested phases in a phase-segregated system [37,38]. In this measurement protocol, the sample is initially cooled from 300 K to 4 K under various applied magnetic fields, $H_a$. Subsequently, the magnetic field is switched from $H_a$ to a fixed measuring field ($H_m = 0.5$ T) at 4 K. Further, the warming data is recorded from 4 K to 300 K in $H_m$. Figure 7 (b) highlights the results of the CHUF measurements. From these measurements, it can be seen that the insulating state is not affected by the CHUF measurement protocol. However, the metallic state below 70 K is highly modified by the value of $H_a$ used for cooling the sample. For T ≤ 70 K, in the metallic state, the same $H_m$ gives different curves for different $H_a$. To understand the thermal evolution of MR due to CHUF in the metallic region of the heterostructure, the MR ratio (%) is calculated using equation (1). The data is presented in the inset of figure 7 (b), where R(0) and R(H) are the resistance values recorded in the zero-field cooled warming cycle, and $H_a$ (= 0 T, 1 T, 3 T, 5 T, 10 T, and 15 T) cooled $H_m$ (= 0.5 T) warming cycle, respectively. The competition between the two antagonistic properties in the junction region of the heterostructure is evident from the CHUF measurement. Kinetic arrest is possible in this regime, which means that the superconducting grains and the spin-polarized ferromagnetic domains coexist at low temperatures. With the increasing value of $H_a$, the scattering centres present below ~ 40 K are destroyed, thereby exhibiting a huge negative CMR of nearly 90 % at 15 T near low temperatures.

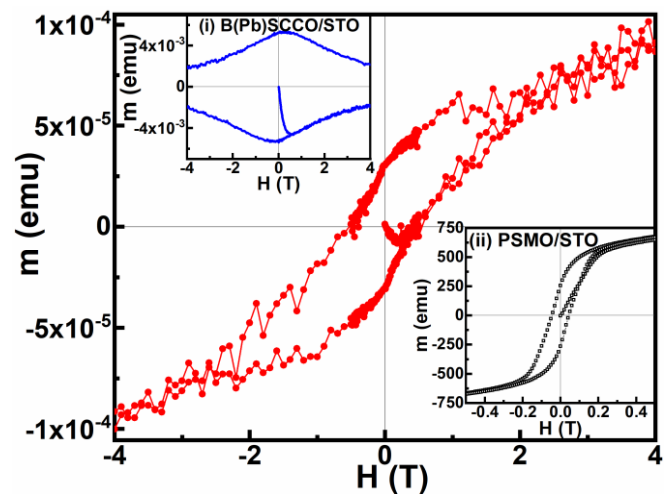

*Figure 8. Magnetic-field dependent magnetic moment (m-H) plot of B(Pb)SCCO/PSMO/STO heterostructure at 10 K. Inset (i) and (ii) represents the m-H plot of the single layers of B(Pb)SCCO/STO and PSMO/STO, respectively.*

The magnetic property of the B(Pb)SCCO/PSMO/STO heterostructure was determined using the magnetization measurements. Figure 8 represents the magnetic field-dependent magnetic moment (m-H) plot obtained for the heterostructure for a representative temperature, T = 10 K. The m-H data obtained for their single layer B(Pb)SCCO and PSMO thin films are also plotted in Figure 8 as insets (i) and (ii), respectively, for comparison. The m-H plot obtained for the heterostructure shows superimposed ferromagnetic and superconducting behaviour.

### *PSMO/B(Pb)SCCO/STO*

Structural properties

The XRD pattern obtained for the fabricated PSMO/B(Pb)SCCO/STO heterostructure in the powder X-ray diffraction (XRD) mode and high-resolution X-ray diffraction (HR-XRD) mode near the *(2 0 0)* reflection of STO are shown in figure 9 and the corresponding inset, respectively. The presence of the individual layers in the heterostructure with *c*-axis preferential orientation was confirmed from the XRD pattern. A broad peak pertaining to the PSMO layer of the heterostructure was obtained at 24.96° near the *(2 0 0)* reflection of STO, as shown in the inset of Figure 9. The *c*-lattice parameter of PSMO, as determined by using Bragg's law was found to be ~ 7.65 Å and ~ 7.64 Å in the single layer and the heterostructure, respectively, which are comparable to the corresponding PSMO bulk value of ~ 7.67 Å, as mentioned earlier.

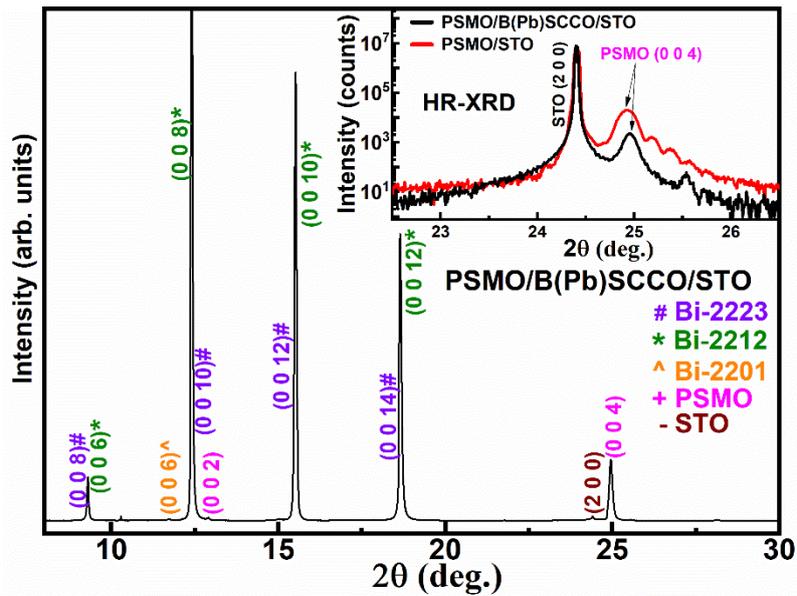

*Figure 9. XRD pattern obtained for the PSMO/B(Pb)SCCO/STO heterostructure thin film. Inset shows the HR-XRD of PSMO/B(Pb)SCCO/STO heterostructure and PSMO/STO single layer near (2 0 0) reflection of STO.*

Morphological analyses

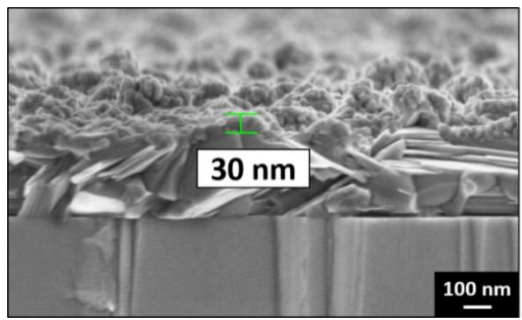

*Figure 10. Cross-sectional SEM of the PSMO/B(Pb)SCCO/STO heterostructure illustrating the individual layers and their thicknesses.*

The thickness of the grown PSMO/B(Pb)SCCO/STO heterostructure was determined using the fracture cross-sectional view mode of SEM. Figure 10 represents the cross-sectional SEM micrograph of the PSMO/B(Pb)SCCO/STO heterostructure. The micrograph shows the presence of PSMO layer of ~ 30 nm grown on top of ~ 300 nm layered B(Pb)SCCO deposited on the STO substrate. However, the interface is not smooth since the B(Pb)SCCO layer is rough. The presence of all the elements in the heterostructure was confirmed elemental analysis using energy-dispersive X-ray spectroscopy (EDS).

Magnetotransport and magnetization measurements

Figure 11 (a) shows the temperature-dependent resistance (R(T)) plot obtained for the PSMO/B(Pb)SCCO/STO heterostructure for H = 0 T, in the first regime, i.e., on the B(Pb)SCCO side, where $V_1$ is measured between the leads 2 and 3. In this regime, a two-step transition is observed, with the first transition having a $T_{C,ON1}$ at 107 K, and the second transition $T_{C,ON2}$ at 74 K. $T_{C,ON1}$ and $T_{C,ON2}$ can be attributed to the superconducting drops corresponding to Bi-2223 and Bi-2212 phase, whose values may be suppressed due to the proximity effect. $T_{C,OFF}$ in this regime is found to be at 65 K. However, on the single layer of B(Pb)SCCO, the two-step transition is not prevalent, and instead, there is a single narrow transition with $T_{C,ON}$ at 113 K and $T_{C,OFF}$ at 107 K [28], as shown in the inset of figure 5 (a). Despite the presence of the two steps in the resistive transition on the B(Pb)SCCO side, the zero-resistance state is intact, as shown in Figure 11 (a). Figure 11 (b) shows the R(T) plot obtained for the PSMO/B(Pb)SCCO/STO heterostructure for H = 0 T in the second regime, i.e., across the junction between the B(Pb)SCCO layer and the PSMO/B(Pb)SCCO/STO heterostructure, where the voltage drop $V_2$ is measured between the electrical leads 4 and 5. Across the junction, the trend of the resistive curve is semiconducting from room temperature down to a temperature of 74 K, and thereafter, it shows a metallic trend with a drop in the resistance value by one order till 50 K. The resistance starts increasing on further decreasing the temperature, showing an insulating behaviour. Figure 11 (c) shows the R(T) curve obtained for the PSMO/B(Pb)SCCO/STO heterostructure for H = 0 T in the third regime, i.e., on the PSMO surface of the PSMO/B(Pb)SCCO/STO heterostructure, where $V_3$ is measured between the leads 6 and 7. It is interesting to note that a superconducting transition is observed on the heterostructure in the third regime, albeit at a lower temperature. In this regime, the $T_{C,ON}$ is observed at 74 K, and the $T_{C,OFF}$ is at 60 K. Figure 11 (d) illustrates the comparative plot of the R(T) behaviour obtained for the PSMO/B(Pb)SCCO/STO heterostructure for the three regimes. The comparative plot shows that the second transition on the B(Pb)SCCO side, the superconducting transition on the PSMO surface of the heterostructure, and the insulating to metallic transition observed across the junction between the B(Pb)SCCO layer and the heterostructure are all concomitant.

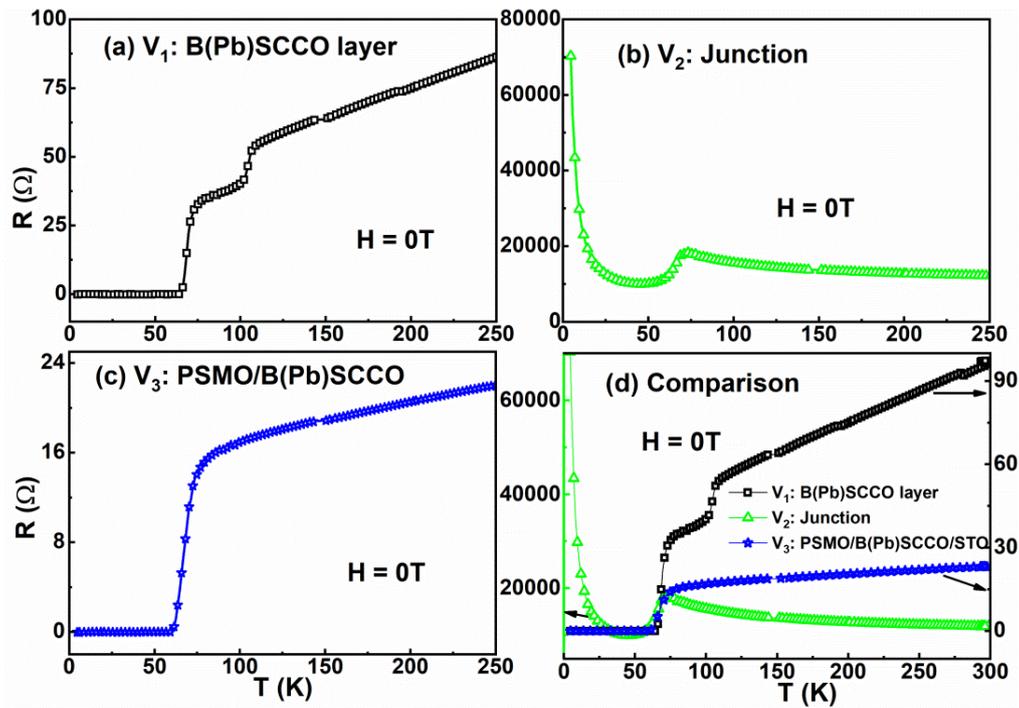

Figure 11. R(T) plots of PSMO/B(Pb)SCCO/STO heterostructure at H=0 T, for resistance measured across (a) $V_1$: B(Pb)SCCO, (b) $V_2$: Junction (c) $V_3$: PSMO/B(Pb)SCCO heterostructure, and (d) a comparative plot for the three regimes

The field dependent R(T) plots obtained from the magnetotransport measurements performed on the B(Pb)SCCO layer and the junction region of the PSMO/B(Pb)SCCO/STO heterostructure for H||c configuration, are depicted in Figure 12 (a) and (b), respectively. The resistance values in the R(T) curves are normalized to its value at 80 K for ease of comparison. It is observed that the resistive curve on the B(Pb)SCCO layer fans out, accompanied by suppression in the $T_{C,OFF}$ with the application of the magnetic field. Interestingly, the resistive curve in the junction regime of the heterostructure also fans out in the metallic region (50 K > T > 80 K) of the curve with the application of H, indicating that despite the suppression in the superconducting property, the condensation energy is still higher than the magnetic exchange energy. The insulating behaviour obtained in this region for T < 50 K can be due to an increase in the scattering centres due to the competition of the spin exchange interaction of the manganite with the Cooper-pairs of the superconductor at the superconductor/manganite interface in the step geometry. Similar results have been reported in the coplanar trilayers of $La_{0.7}Sr_{0.3}MnO_3/Bi_2Sr_2CaCu_2O_8/La_{0.7}Sr_{0.3}MnO_3$ [20]. Hence, in both regimes, below 80 K, the behaviour identical to that of a superconductor is observed, wherein the resistance increases due to the energy dissipation due to the movement of vortices in response to the applied H.

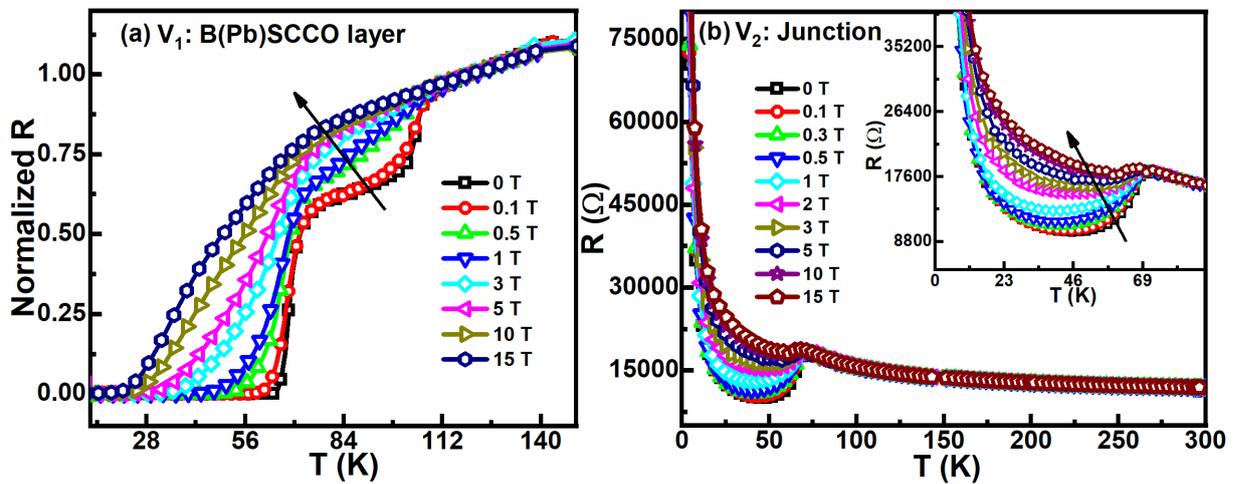

*Figure 12. The magnetic field-dependent normalized R(T) curves of PSMO/B(Pb)SCCO/STO heterostructure for H||c measured across (a) leads 2 and 3 (B(Pb)SCCO layer) and (b) leads 4 and 5 (Junction between the B(Pb)SCCO layer and the heterostructure). Inset in (b) highlights the fanning out of the R(T) curve as a function of H.*

The junction regime in the B(Pb)SCCO/PSMO/STO heterostructure exhibited a huge CMR while the junction regime in the PSMO/B(Pb)SCCO/STO heterostructure exhibited superconducting behaviour. Thus, while the magnetic exchange energy dominates in the former, the superconducting condensation energy dominates in the latter. Thus, the results obtained from the proximity studies between the superconductor and the CMR manganite at the junction regime in the heterostructures investigated reveal that the stacking sequence of the individual layers can tailor the ground state properties at the junction regime.

The temperature-dependent normalized resistance R(T) plots obtained from the magnetotransport measurements performed on the PSMO/B(Pb)SCCO/STO heterostructure for H∥$c$ configuration, corresponding to $V_3$ measured on PSMO/B(Pb)SCCO/STO heterostructure, are depicted in Figure 13 (a). The resistance in these curves is also normalized to its value at 80 K for ease of comparison. The inset (i) of Figure 13 (a) depicts the determination of $T_{C,ON}$ by tangent intersection method. At H = 0 T, $T_{C,ON}$ is obtained at 74 K, and $T_{C,OFF}$ is at 60 K with a superconducting transition width ($\Delta T_C$) of 16 K. The $T_{MIT}$ in an annealed 30 nm PSMO thin film was at ~ 170 K, as discussed earlier [36]. Hence, the top layer of PSMO is in the ferromagnetic metallic state when the bottom B(Pb)SCCO layer is superconducting. Nonetheless, the competition between the two antagonistic properties has resulted in inducing superconductivity on the PSMO surface of the heterostructure with a reduced $T_C$. The coherence length of B(Pb)SCCO, as reported earlier [28,30], is nearly 1.34 nm. Hence, this long-range proximity-induced superconductivity on the ferromagnetic PSMO can be explained only if we consider the spin flipping of some of the Cooper pairs. Since the thickness of the magnetic layer is only a few tens of nm, it is possible to induce superconductivity in the ferromagnet by spin flipping [39,40].

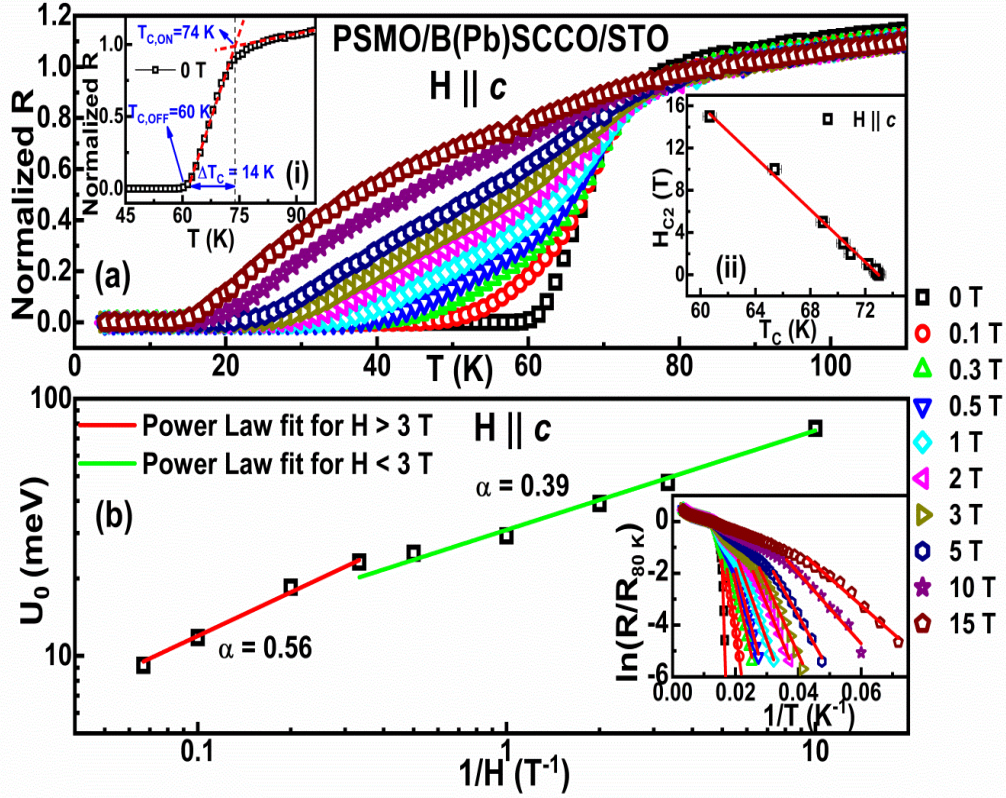

*Figure 13. (a) The magnetic field-dependent normalized R(T) curves of PSMO/B(Pb)SCCO/STO heterostructure for H||c measured across lead 6 and 7. Inset (i) and (ii) in (a) illustrates the determination of $T_{C,ON}$, and the variation of $H_{C2}$ as a function of $T_C$ for the heterostructure for H||c, respectively. (b) Determination of $U_0$ from Arrhenius plot (inset) and its power-law dependence on H in log-log scale for H||c.*

The observed reduction in the value of $T_C$ on the heterostructure can be due to the following reasons: the reduction in the number density of Cooper-pairs in the B(Pb)SCCO due to its leakage into FM PSMO and the diffusion and transport of the spin-polarized electrons from the PSMO into superconducting B(Pb)SCCO. The broadened resistive curve with a $\Delta T_C \sim 16$ K, obtained in the heterostructure compared to the single layer ($\Delta T_C \sim 6$ K), can be due to the inherent magnetic exchange interaction of the FM PSMO. The heterostructure exhibits metallic behaviour above $T_{C,ON}$. However, near the $T_{C,OFF}$, broadening of the resistive curve due to energy dissipation by the thermally activated vortex movement is observed as a response to applied H [41]. This regime can be explained by the

Arrhenius relation given by equation (2), where U(T, H) is the activation energy required for the bundle of vortices to hop from a pinning site of potential $U_0$ to another and $k_B$ is the Boltzmann constant [42]. The inset in Figure 13 (b) represents the determination of the value of $U_0$ from the plot of $\ln(R/R_0)$ as a function of $T^{-1}$ for $H \| c$. The slope value near the linear region of the plot estimates the value of $U_0$. As perceived from Figure 13 (b), the estimated $U_0$ is found to have a power-law dependence of the form $U_0 \sim H^{-\alpha}$ on H with an α value of 0.56 for H ≥ 3 T, indicating a 2D pancake arrangement of the vortices having a double kink in the vortex lattice line in the superconducting matrix [43–45]. However, the data point corresponding to H < 3 T deviates from this linear behaviour and scales H with an α value of 0.39. This may be because of the single vortex pinning in the low field regime (H < 3 T), as the applied magnetic field may not be sufficient to establish collective pinning in this regime.

$$\rho = \rho_0 \exp\left(-U(T,H)/k_B T\right) \qquad (2)$$

The upper critical field ($H_{C2}(0)$) of the heterostructure was determined using the reduced Werthamer-Helfand-Hohenberg (WHH) relation [46] given by equation (3) by adopting 80 % criteria. Inset (ii) of Figure 13 (a) represents the WHH plot obtained for the PSMO/B(Pb)SCCO/STO heterostructure for the $H \| c$ configuration. The value of $H_{C2}(0)$ thus obtained is 63 T, which is found to be drastically suppressed compared to the B(Pb)SCCO single layer films. This drastic suppression of $H_{C2}(0)$ can be attributed to the magnetic flux penetration due to the internal magnetic exchange interaction of PSMO into the superconducting matrix of B(Pb)SCCO. The coherence length value $\xi_{ab}$, calculated using the Ginzburg-Landau relation given by equation (4), is found to be 2.29 nm. Interestingly, the superconducting critical properties of B(Pb)SCCO are more adversely affected in proximity with the ferromagnetic PSMO than that to the half-doped charge-ordered manganite $Pr_{0.5}Ca_{0.5}MO_3$ [24]. The larger suppression in the superconducting critical parameters ($T_C$ and

$H_{C2}$) of B(Pb)SCCO in proximity with PSMO compared to $Pr_{0.5}Ca_{0.5}MO_3$ is due to the larger magnetic exchange energy of the ferromagnetic metallic PSMO.

$$H_{C2}(0) = -0.693 T_C \left( dH_{C2}/dT \right)_{T_C} \tag{3}$$

$$\mu_0 H_C^{\|c}(0) = \left( \frac{\phi_0}{2\pi \xi_{ab}^2(0)} \right) \tag{4}$$

The magnetic property of the PSMO/B(Pb)SCCO/STO heterostructure was determined using the magnetization measurements. Figure 14 represents the magnetic field-dependent magnetic moment (m-H) plot obtained for the heterostructure for T = 50 K, 77 K, and 100 K. The m-H plot shows superimposed ferromagnetic and superconducting behaviour. At 100 K, the m-H resembles the ferromagnetic hysteresis loop. This confirms the presence of the ferromagnetic order in the heterostructure.

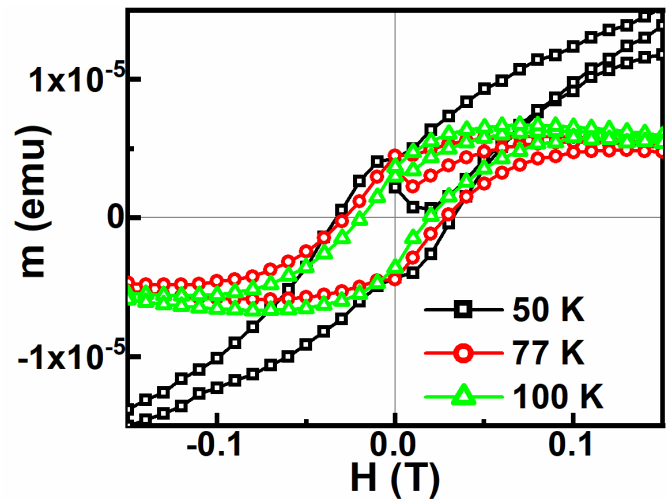

*Figure 14. Magnetic field dependent magnetic moment (m-H) plot of PSMO/B(Pb)SCCO/STO heterostructure.*

**Conclusions**

From the above studies on the heterostructures of B(Pb)SCCO and PSMO, it is surmised that the proximity of the thin magnetic layer of PSMO can suppress the superconductivity in B(Pb)SCCO drastically. The reduction in the superconducting critical parameters is attributed to the combined detrimental effect of the magnetic exchange interaction of PSMO, the leakage of Cooper pairs from B(Pb)SCCO, and the diffusion and transport of spin-polarized electrons into the superconducting matrix of B(Pb)SCCO from the ferromagnetic metallic PSMO. As a consequence of the proximity effect, the huge CMR ratio as high as ~ 99 % observed in the heterostructure, can make the thin film heterostructures promising candidates for potential technological applications. Furthermore, the results obtained from the proximity studies between the superconductor and the CMR manganite at the junction regime in the heterostructures reveal that the stacking sequence of the individual layers can tailor the ground state properties at the junction regime.

**Acknowledgement**

The authors thank the Indus-2 Synchrotron facility, RRCAT, Indore (India) for the XRD measurements. The authors acknowledge UGC-DAE-CSR, Kalpakkam Node (India), for the SEM-EDS, magnetotransport, and SQUID-VSM facilities. Gayathri V is grateful to the Department of Atomic energy, Government of India, for the financial support during her research work.

**References**

(1) Linder, J.; Robinson, J. W. A. Superconducting Spintronics. *Nat. Phys.* **2015**, *11* (4), 307–315. https://doi.org/10.1038/nphys3242.

(2) Buzdin, A. I. Proximity Effects in Superconductor-Ferromagnet Heterostructures. *Rev. Mod. Phys.* **2005**, *77* (3), 935–976. https://doi.org/10.1103/RevModPhys.77.935.

(3) Bergeret, F. S.; Volkov, A. F.; Efetov, K. B. Odd Triplet Superconductivity and Related Phenomena in Superconductor-Ferromagnet Structures. *Rev. Mod. Phys.* **2005**, *77* (4), 1321–1373. https://doi.org/10.1103/RevModPhys.77.1321.

(4) De Gennes, P. G. Boundary Effects in Superconductors. *Rev. Mod. Phys.* **1964**, *36* (1), 225–237.

(5) Parker, C. V.; Pushp, A.; Pasupathy, A. N.; Gomes, K. K.; Wen, J.; Xu, Z.; Ono, S.;

Gu, G.; Yazdani, A. Nanoscale Proximity Effect in the High-Temperature Superconductor $Bi_2Sr_2CaCu_2O_{8+\delta}$ Using a Scanning Tunneling Microscope. *Phys. Rev. Lett.* **2010**, *104* (11), 1–4. https://doi.org/10.1103/PhysRevLett.104.117001.

(6) Cherkez, V.; Cuevas, J. C.; Brun, C.; Cren, T.; Ménard, G.; Debontridder, F.; Stolyarov, V. S.; Roditchev, D. Proximity Effect between Two Superconductors Spatially Resolved by Scanning Tunneling Spectroscopy. *Phys. Rev. X* **2014**, *4* (1), 11033. https://doi.org/10.1103/PhysRevX.4.011033.

(7) Gayathri, V.; Sathyanarayana, A. T.; Amaladass, E. P.; Vinod, K.; Kumary, T. G.; Pandian, R.; Mani, A. Evolution of Superconducting Properties of Coexistent Bi-2212 and Bi-2223 Phases in BSCCO. *Indian J. pure Appl. Phys.* **2021**, *59* (May), 391–397. https://doi.org/10.1038/s41563-019-0476-3.

(8) Jiang, J. S.; Davidovic, D.; Reich, D. H.; Chien, C. L. Oscillatory Superconducting Transition Temperature in Nb/Gd Multilayers. *Phys. Rev. Lett.* **1995**, *74* (2), 314–317.

(9) Bourgeois, O.; Dynes, R. C. Strong Coupled Superconductor in Proximity with a Quench-Condensed Ferromagnetic Ni Film: A Search for Oscillating Tc. *Phys. Rev. B - Condens. Matter Mater. Phys.* **2002**, *65* (14), 1445031–1445035. https://doi.org/10.1103/PhysRevB.65.144503.

(10) Mühge, T.; Theis-Bröhl, K.; Westerholt, K.; Zabel, H.; Garif'yanov, N. N.; Goryunov, Y. V; Garifullin, I. A.; Khaliullin, G. G. Influence of Magnetism on Superconductivity in Epitaxial Fe/Nb Bilayer Systems. *Phys. Rev. B* **1998**, *57* (9), 5071–5074.

(11) Aarts, J.; Geers, J. M. E.; Brück, E.; Golubov, A. A.; Coehoorn, R. Interface Transparency of Superconductor/Ferromagnetic Multilayers. *Phys. Rev. B - Condens. Matter Mater. Phys.* **1997**, *56* (5), 2779–2787. https://doi.org/10.1103/PhysRevB.56.2779.

(12) Karelina, L. N.; Hovhannisyan, R. A.; Golovchanskiy, I. A.; Chichkov, V. I.; Ben Hamida, A.; Stolyarov, V. S.; Uspenskaya, L. S.; Erkenov, S. A.; Bolginov, V. V; Ryazanov, V. V. Scalable Memory Elements Based on Rectangular SIsFS Junctions. *J. Appl. Phys.* **2021**, *130* (17), 173901. https://doi.org/10.1063/5.0063274.

(13) Mani, A.; Kumary, T. G.; Hsu, D.; Lin, J. G.; Chern, C.-H. Modulation of Superconductivity by Spin Canting in a Hybrid Antiferromagnet/Superconductor

Oxide. *Appl. Phys. Lett.* **2009**, *94* (7), 072509. https://doi.org/10.1063/1.3087000.

(14) Mani, A.; Kumary, T. G.; Hsu, D.; Lin, J. G. Current and Field Dependent Proximity Effects in the $Nd_{0.43}Sr_{0.57}MnO_3/YBa_2Cu_3O_7$ Heterostructure. *J. Appl. Phys.* **2009**, *105* (10), 103915. https://doi.org/10.1063/1.3130605.

(15) Mani, A.; Kumary, T. G.; Hsu, D.; Lin, J. G. Strain Enhanced Spin Polarization in $Nd_{0.43}Sr_{0.57}MnO_3/YBa_2Cu_3O_7$ Bilayers. *J. Appl. Phys.* **2008**, *104* (5), 053910. https://doi.org/10.1063/1.2976365.

(16) Mani, A.; Geetha Kumary, T.; Lin, J. G. Thickness Controlled Proximity Effects in C-Type Antiferromagnet/Superconductor Heterostructure. *Sci. Rep.* **2015**, *5* (1), 12780. https://doi.org/10.1038/srep12780.

(17) Kumari, S.; Anas, M.; Raghav, D. S.; Chauhan, S.; Siwach, P. K.; Malik, V. K.; Singh, H. K. A Comparative Study of Superconductivity and Thermally Activated Flux Flow of $YBa_2Cu_3O_{7-\delta}$ and $YBa_2Cu_3O_{7-\delta}/La_{1-x-y}Pr_xCa_yMnO_3$ Bilayers. *J. Supercond. Nov. Magn. 2022 3511* **2022**, *35* (11), 3225–3240. https://doi.org/10.1007/S10948-022-06381-8.

(18) Ryazanov, V. V; Oboznov, V. A.; Rusanov, A. Y.; Veretennikov, A. V; Golubov, A. A.; Aarts, J. Coupling of Two Superconductors through a Ferromagnet: Evidence for a π Junction. *Phys. Rev. Lett.* **2001**, *86* (11), 2427–2430. https://doi.org/10.1103/PhysRevLett.86.2427.

(19) Johnsen, L. G.; Jacobsen, S. H.; Linder, J. Magnetic Control of Superconducting Heterostructures Using Compensated Antiferromagnets. *Phys. Rev. B* **2021**, *103* (6), L060505-. https://doi.org/10.1103/PhysRevB.103.L060505.

(20) Joseph, D. P.; Lin, J. G. Non-Local Spin Injection Effects in Coplanar $La_{0.7}Sr_{0.3}MnO_3/Bi_2Sr_2CaCu_2O_8/La_{0.7}Sr_{0.3}MnO_3$ Tri-Layer. *AIP Conf. Proc.* **2015**, *1665* (1), 130001. https://doi.org/10.1063/1.4918149.

(21) Daniel, P. J.; Lin, J. G. Investigation of Optimal Growth Conditions of $La_{0.7}Sr_{0.3}MnO_3-Bi_2Sr_2Ca_1Cu_2O_8$ Heterostructures. *J. Am. Ceram. Soc.* **2013**, *96* (2), 481–484. https://doi.org/10.1111/jace.12056.

(22) Paredes, O.; Baca, E.; Fuchs, D.; Morán, O. Crossover from Negative to Positive Magnetoresistance in Superconductor/Ferromagnet Composites Thick Films. *Phys. C*

*Supercond.* **2010**, *470* (21), 1911–1915. https://doi.org/https://doi.org/10.1016/j.physc.2010.06.012.

(23) Krivoruchko, V. N.; Tarenkov, V. Y. Percolation Transitions in D-Wave Superconductor-Half-Metallic Ferromagnet Nanocomposites. *Low Temp. Phys.* **2019**, *45* (5), 476. https://doi.org/10.1063/1.5097355.

(24) Gayathri, V.; Amaladass, E. P.; Vinod, K.; Sathyanarayana, A. T.; Geetha Kumary, T.; Mani, A. Variation in the Superconducting Properties of $Bi_{1.75}Pb_{0.25}Sr_2Ca_2Cu_3O_{10+\delta}$ in Proximity with $Pr_{0.5}Ca_{0.5}MnO_3$. *Cryogenics (Guildf).* **2022**, *126*, 103542. https://doi.org/10.1016/j.cryogenics.2022.103542.

(25) Bhatt, H.; Kumar, Y.; Prajapat, C. L.; Kinane, C. J.; Caruana, A.; Langridge, S.; Basu, S.; Singh, S. Correlation of Magnetic and Superconducting Properties with the Strength of the Magnetic Proximity Effect in $La_{0.67}Sr_{0.33}MnO_3/SrTiO_3/YBa_2Cu_3O_{7-\delta}$ Heterostructures. *ACS Appl. Mater. Interfaces* **2022**, *14* (6), 8565–8574. https://doi.org/10.1021/acsami.1c22676.

(26) Geetha Kumary, T.; Amaladass, E. P.; Nithya, R.; Mani, A. Strain-Enhanced Colossal Magnetoresistance in $Pr_{0.6}Sr_{0.4}MnO_3$ Thin Films. *J. Supercond. Nov. Magn.* **2016**, *29* (10), 2685–2690. https://doi.org/10.1007/s10948-016-3590-3.

(27) Gayathri, V.; Sathyanarayana, A. T.; Shukla, B.; Geetha Kumary, T.; Kumar, S.; Chandra, S.; Mani, A. High-Pressure Studies on Pristine and Pb-Substituted Bi-Based High-Temperature Superconductor. *Bull. Mater. Sci.* **2022**, *45* (4), 207. https://doi.org/10.1007/s12034-022-02797-zS.

(28) Gayathri, V.; Amaladass, E. P.; Geetha Kumary, T.; Mani, A. Tuning of Anisotropy in Bi-Based High Temperature Superconducting Thin Films. *Thin Solid Films* **2022**, *757* (July), 139418. https://doi.org/10.1016/j.tsf.2022.139418.

(29) Gayathri, V.; Sathyanarayana, A. T.; Vinod, K.; Geetha Kumary, T.; Mani, A. Microstructure and Substrate Dependent Enhanced Critical Current Density in Pb-Substituted Bi-Based High Temperature Superconducting Thin Films. *Phys. Scr.* **2022**, *97* (12), 125823. https://doi.org/10.1088/1402-4896/aca12b.

(30) Gayathri, V.; Bera, S.; Amaladass, E. P.; Geetha Kumary, T.; Pandian, R.; Mani, A. Effects of Pb Assisted Cation Chemistry on the Superconductivity of BSCCO Thin

Films. *Phys. Chem. Chem. Phys.* **2021**, *23* (22), 12822–12833. https://doi.org/10.1039/d1cp01262b.

(31) Gayathri, V.; Kumary, T. G.; Amaladass, E. P.; Sathyanarayana, A. T.; Mani, A. Studies on Colossal Magnetoresistance Behaviour of $Pr_{0.6}Sr_{0.4}MnO_3/Pr_{0.5}Ca_{0.5}MnO_3$ Heterostructure Films. *J. Supercond. Nov. Magn.* **2021**, *34* (7), 1955–1960. https://doi.org/10.1007/s10948-021-05889-9.

(32) Gupta, P.; Rao, P. N.; Swami, M. K.; Bhakar, A.; Lal, S.; Garg, S. R.; Garg, C. K.; Gauttam, P. K.; Kane, S. R.; Raghuwanshi, V. K.; Rai, S. K. BL-02: A Versatile X-Ray Scattering and Diffraction Beamline for Engineering Applications at Indus-2 Synchrotron Source. *J. Synchrotron Radiat.* **2021**, *28*, 1193–1201. https://doi.org/10.1107/S1600577521004690.

(33) Koyama, S.; Endo, U.; Kawai, T. Preparation of Single 110 K Phase of the Bi-Pb-Sr-Ca-Cu-O Superconductor. *Jpn. J. Appl. Phys.* **1988**, *27* (Part 2, No. 10), L1861–L1863. https://doi.org/10.1143/jjap.27.l1861.

(34) Jin, S.-G.; Zhu, Z.-Z.; Liu, L.-M.; Huang, Y.-L. Water Reactions of Superconducting $Bi_2Sr_2CaCu_2O_8$ Phase at O°C and Ambient Temperature. *Solid State Commun.* **1990**, *74* (10), 1087–1090. https://doi.org/https://doi.org/10.1016/0038-1098(90)90715-N.

(35) Boujelben, W.; Ellouze, M.; Cheikh-Rouhou, A.; Pierre, J.; Cai, Q.; Yelon, W. B.; Shimizu, K.; Dubourdieu, C. Neutron Diffraction, NMR and Magneto-Transport Properties in the $Pr_{0.6}Sr_{0.4}MnO_3$ Perovskite Manganite. *J. Alloys Compd.* **2002**, *334* (1), 1–8.

(36) Gayathri, V.; Amaladass, E. P.; Sathyanarayana, A. T.; Geetha Kumary, T.; Pandian, R.; Gupta, P.; Rai, S. K.; Mani, A. Interfacial Interaction Driven Enhancement in the Colossal Magnetoresistance Property of Ultra-Thin Heterostructure of $Pr_{0.6}Sr_{0.4}MnO_3$ in Proximity with $Pr_{0.5}Ca_{0.5}MnO_3$. *Sci. Rep.* **2023**, *13* (1), 2315. https://doi.org/10.1038/s41598-023-28314-8.

(37) Banerjee, A.; Kumar, K.; Chaddah, P. Conversion of a Glassy Antiferromagnetic-Insulating Phase to an Equilibrium Ferromagnetic-Metallic Phase by Devitrification and Recrystallization in Al Substituted $Pr_{0.5}Ca_{0.5}MnO_3$. *J. Phys. Condens. Matter* **2009**, *21* (2), 026002. https://doi.org/10.1088/0953-8984/21/2/026002.

(38) Roy, S. B.; Chaddah, P. Phase-Coexistence and Glass-like Behavior in Magnetic and Dielectric Solids with Long-Range Order. *Phys. status solidi* **2014**, *251* (10), 2010–2018. https://doi.org/10.1002/pssb.201451004.

(39) Yao, X.; Jin, Y.; Li, M.; Li, Z.; Cao, G.; Cao, S.; Zhang, J. Coexistence of Superconductivity and Ferromagnetism in La$_{1.85}$Sr$_{0.15}$CuO$_4$-La$_{2/3}$Sr$_{1/3}$MnO$_3$ Matrix Composites. *J. Alloys Compd.* **2011**, *509* (18), 5472–5476. https://doi.org/10.1016/j.jallcom.2011.02.126.

(40) Cai, R.; Yunyan, Y.; Lv, P.; Ma, Y.; Xing, W.; Li, B.; Ji, Y.; Zhou, H.; Shen, C.; Jia, S.; Xie, X.; Žutić, I.; Sun, Q.-F.; Han, W. Evidence for Anisotropic Spin-Triplet Andreev Reflection at the 2D van Der Waals Ferromagnet/Superconductor Interface. *Nat. Commun.* **2021**, *12*. https://doi.org/10.1038/s41467-021-27041-w.

(41) Kucera, J. T.; Orlando, T. P.; Virshup, G.; Eckstein, J. N. Magnetic-Field and Temperature Dependence of the Thermally Activated Dissipation in Thin Films of Bi$_2$Sr$_2$CaCu$_2$O$_{8+\delta}$. *Phys. Rev. B* **1992**, *46* (17), 11004–11013. https://doi.org/10.1103/PhysRevB.46.11004.

(42) Kundu, H. K.; Amin, K. R.; Jesudasan, J.; Raychaudhuri, P.; Mukerjee, S.; Bid, A. Effect of Dimensionality on the Vortex Dynamics in a Type-II Superconductor. *Phys. Rev. B* **2019**, *100* (17), 174501. https://doi.org/10.1103/PhysRevB.100.174501.

(43) Yamasaki, H.; Endo, K.; Kosaka, S.; Umeda, M.; Yoshida, S.; Kajimura, K. Scaling of the Flux Pinning Force in Epitaxial Bi$_2$Sr$_2$Ca$_2$Cu$_3$O$_x$ Thin Films. *Phys. Rev. Lett.* **1993**, *70* (21), 3331–3334. https://doi.org/10.1103/PhysRevLett.70.3331.

(44) Clem, J. R. Two-Dimensional Vortices in a Stack of Thin Superconducting Films: A Model for High-Temperature Superconducting Multilayers. *Phys. Rev. B* **1991**, *43* (10), 7837–7846. https://doi.org/10.1103/PhysRevB.43.7837.

(45) Geshkenbein, V.; Larkin, A.; Feigel'man, M.; Vinokur, V. Flux Pinning and Creep in High-Tc Superconductors. *Phys. C Supercond. its Appl.* **1989**, *162–164*, 239–240. https://doi.org/https://doi.org/10.1016/0921-4534(89)91006-X.

(46) Werthamer, N. R.; Helfand, E.; Hohenberg, P. C. Temperature and Purity Dependence of the Superconducting Critical Field, Hc2. III. Electron Spin and Spin-Orbit Effects. *Phys. Rev.* **1966**, *147* (1), 295–302. https://doi.org/10.1103/PhysRev.147.295.